\documentclass[aps,prl,reprint]{revtex4-1} 
\usepackage[pdftex]{graphicx}
\usepackage{siunitx}
\usepackage{xspace}
\pdfoutput=1

\newcommand{\dto}{Dy$_{2}$Ti$_{2}$O$_{7}$\xspace}
\newcommand{\yto}{Y$_{2}$Ti$_{2}$O$_{7}$\xspace}

\begin{document}

\title{Critical speeding-up near the  monopole liquid-gas transition \\in magnetoelectric spin-ice}

\author{Christoph P. Grams$^1$} 

\author{Martin Valldor$^{1,3}$}

\author{Markus Garst$^2$}

\author{Joachim Hemberger$^1$}

\affiliation{$^1$II. Physikalisches Institut, Universit\"at zu K\"oln, Z\"ulpicher Str. 77, 50937 K\"oln, Germany}  
\affiliation{$^2$Institut f\"ur Theoretische Physik, Universit\"at zu K\"oln, Z\"ulpicher Str. 77, 50937 K\"oln, Germany}  
\affiliation{$^3$Max Planck Institute for Chemical Physics of Solids, N\"othnitzer Stra\ss{}e 40, 01187 Dresden, Germany}

\begin{abstract}
Competing interactions in the so-called spin-ice compounds stabilize a frustrated ground-state with finite zero-point entropy and, interestingly, emergent magnetic monopole excitations. The properties of these monopoles are at the focus of recent research with particular emphasis on their quantum dynamics. It is predicted that each monopole also possesses an electric dipole moment, 
which allows to investigate their dynamics via the dielectric function $\varepsilon(\nu)$. Here, we report on broadband spectroscopic measurements of $\varepsilon(\nu)$ in \dto down to temperatures of 200\,mK with a specific focus on the
critical endpoint  present for a magnetic field along 
the crystallographic [111] direction. 
Clear critical signatures are revealed in the dielectric response 
when, similarly as in the liquid-gas transition, the density of monopoles changes in a critical manner.
Surprisingly, the dielectric relaxation time $\tau$ exhibits a critical speeding-up  with a significant enhancement of $1/\tau$ as the temperature is lowered towards the critical temperature. 
Besides demonstrating the magnetoelectric character of the emergent monopole excitations, our results reveal unique critical dynamics near the monopole condensation transition.
\end{abstract}

\maketitle

\begin{figure}
\centering
\includegraphics[width=\columnwidth]{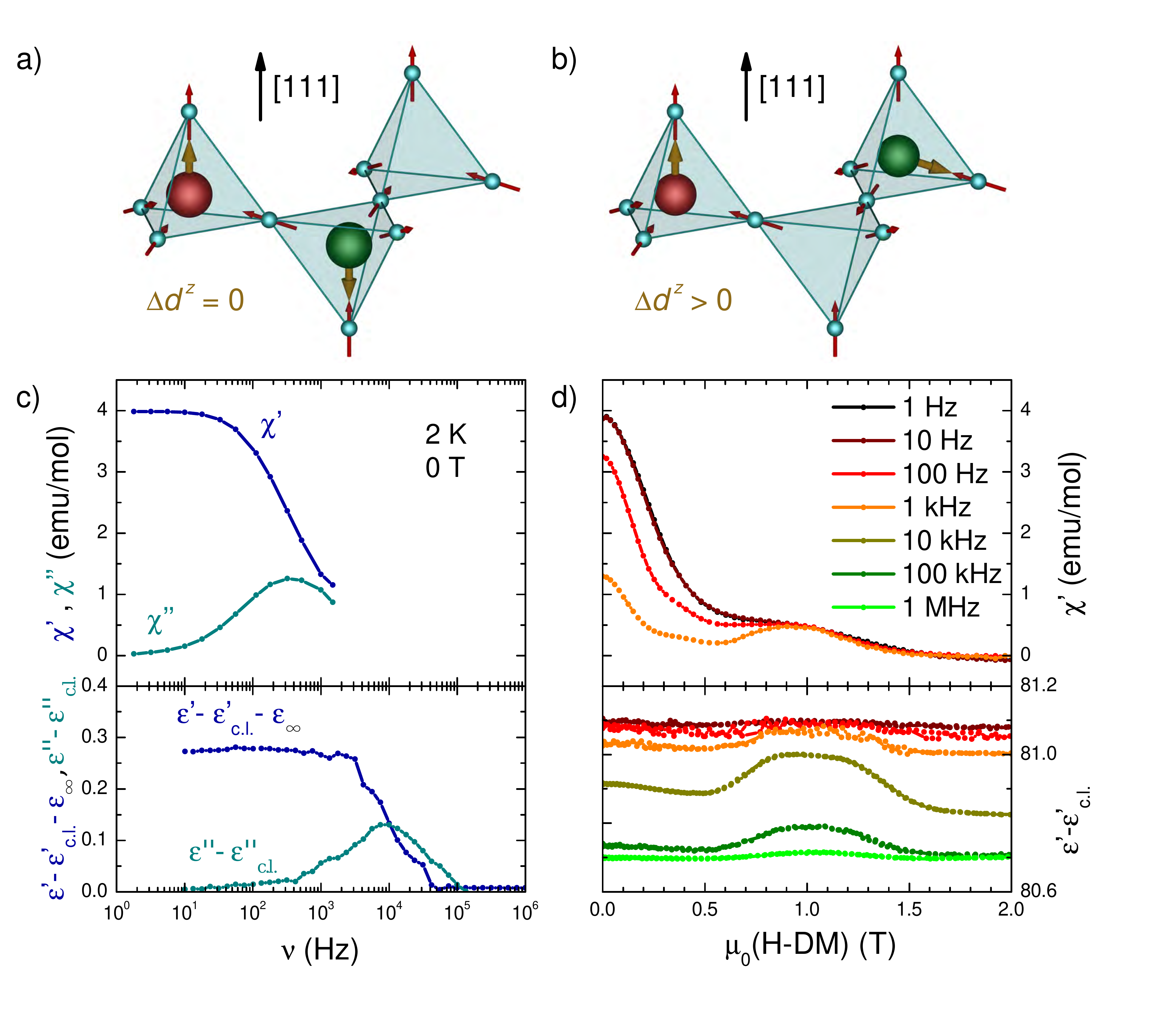}
\caption{(a) Corner-sharing network of tedrahedra in spin-ice with a Ising spin (red arrow) located on each corner. Breaking of the ice rules results in an effective magnetic monopole/anti-monopole pair (red and green balls) each of which carries an electric dipole moment (brown arrow). 
(b) The hopping of a monopole corresponds to flipping a single Ising spin, here giving rise to a non-vanishing net electric dipole moment $\Delta d^z$. 
(c) Spectra of the complex magnetic ac-susceptibility $\chi(\nu)$ (upper frame) and of the complex permittivity $\varepsilon(\nu)$ (lower frame) at 2\,K in zero magnetic field. 
The dielectric function was corrected for a constant loss $\varepsilon_{\rm c.l.}(\nu)$ and a background contribution $\varepsilon_{\infty}$ as described in the text. 
(d) The real parts of magnetic ac-susceptibility (upper frame) and permittivity (lower frame) as a function of internal magnetic field, $H_{i} = H - D M$, along the [111] axis at $T = 2$\,K measured for frequencies between 1\,Hz and 1\,MHz. 
} 
\label{fig:chi1-eps1}
\end{figure}


The magnetic rare earth ions of the spin-ice compounds like \dto form a lattice of corner-sharing tetrahedra. The strong influence of crystal fields forces the magnetic moments to point along the axis connecting the centers of adjacent tetrahedra giving rise to a description in terms of effective Ising spins. In zero magnetic field, the ground state is characterized by a 2-in/2-out spin state configuration, i.e., there are two spins for each tetrahedron that point inwards and two that point outwards. 
The ground state, however, is not unique as there is a multitude of possible realizations of this 2-in/2-out arrangement giving rise to a finite zero-point entropy. 
Given the resemblance to the proton configurations of water-ice, this peculiar magnetism is known as spin-ice \cite{Harris1997,Ramirez1999,Bramwell2001}. 
Flipping a single spin breaks the 2-in/2-out ice rule for two adjacent tetrahedra: the resulting magnetic dipole excitation decomposes into an emergent magnetic monopole and anti-monopole pair that interact via a long-range magnetic Coulomb law \cite{Castelnovo2008,Ryzhkin2005}. Spin-ice thus offers a rare example of fractionalized excitations in a three-dimensional material.

Recently, it has been pointed out by D. Khomskii \cite{Khomskii2012} that each magnetic monopole is also accompanied by an electric dipole. The presence of a magnetic monopole on a tetrahedron results in an electric dipole moment either due to a spontaneous charge redistribution or due to a magnetoelastic coupling. This dipole moment always points along the corner of the tetrahedron that is occupied by the spin which is singled out by the 1-in/3-out or the 1-out/3-in configuration, see Fig.~\ref{fig:chi1-eps1}(a) and (b). 
Furthermore, the hopping of monopoles from one tetrahedron to another is accompanied by a change of the net electric dipole moment. The dynamics of monopoles within the material thus leads to characteristic fluctuations of the electric polarization that, in principle, can be investigated with the help of the dielectric function $\varepsilon(\nu)$. 
This is remarkable as it would provide means to probe explicitly the magnetic monopole excitations in spin-ice. 
Whereas the magnetic response $\chi(\nu)$ is sensitive to both, magnetic monopoles that break the ice rules as well as collective spin-fluctuations that leave them intact, the magnetic part of the dielectric response is mainly susceptible to the former. 
First experimental indications for a magnetoelectric coupling in \dto were reported by Saito {\it et al.} \cite{Saito2005}. 
However, in their work the permittivity was only measured for a fixed frequency of 1\,kHz. In the present work, we use broad-band dielectric spectroscopy and present measurements of $\varepsilon(\nu)$ in \dto for a wide range of frequencies, 1\,Hz\,$ < \nu < $1\,MHz, and temperatures down to 200\,mK.

A comparison between the magnetic and dielectric response in \dto is shown in Fig.~\ref{fig:chi1-eps1}. 
The spin dynamics up to kHz frequencies was already intensively explored by magnetic ac-susceptibility $\chi(\nu)$ measurements \cite{Snyder2001,Matsuhira2011,Yaraskavitch2012,Matthews2012,Bovo2013}. 
In zero magnetic field the corresponding spin relaxation time is found to be temperature independent between 3 and 12\,K but increases exponentially at lower temperatures as the ice rules start to establish themselves. 
A spectrum of $\chi(\nu)$ is shown in the upper frame of Fig.~\ref{fig:chi1-eps1}(c). 
The loss peak in the imaginary part $\chi''(\nu)$ at a temperature of 2\,K is located at $\nu^\chi_p \approx 350$\,Hz, and the resulting relaxation time, $\tau = 1/(2\pi \nu_p)$, is in nice agreement with previously reported results \cite{Matthews2012,Matsuhira2011,Yaraskavitch2012}.
Although our spectral range for $\chi(\nu)$ is also experimentally limited to $\nu\leq 1$\,kHz, it is already apparent that the real part $\chi'(\nu)$ levels off at a finite value at higher frequencies.
One expects that a second relaxation process, which has not been detected yet, eventually leads to a vanishing of $\chi'(\nu)$ at even larger $\nu$. 
Such a two-stage process is common in paramagnetic materials with the slow relaxation process being attributed to the spin-lattice interaction \cite{Snyder2001,Bovo2013}.

A spectrum of the dielectric response $\varepsilon(\nu)$ is shown in the lower frame of Fig.~\ref{fig:chi1-eps1}(c). 
We find that $\varepsilon(\nu)$ is insensitive to the slow relaxation process that governs $\chi(\nu)$. 
Instead, its relaxation dynamics is roughly two orders of magnitude faster with a loss peak in $\varepsilon''(\nu)$ located at roughly $\nu_{p}^\varepsilon \approx 10$\,kHz.
At this point, it is however unclear whether this relaxation process is related to magnetism or rather has a non-magnetic origin.

In order to identify the magnetic contribution to the dielectric response, we investigated its magnetic field dependence. 
Particularly interesting is the application of a magnetic field along the  $[111]$ direction. 
A quarter of all spins then align with the magnetic field at low temperatures due to the Zeeman energy. 
The remaining spins occupy sites of decoupled two-dimensional Kagome lattices. 
The ground state manifold consistent with the ice rules is still degenerate and thus realizes \textit{Kagome-ice}. 
At a critical internal field of $\mu_0 H_{i,c} = 
1$\,T in \dto, \cite{Sakakibara2003,Hiroi2003,Aoki2004,Kolland2012} it becomes advantageous to break the ice rules and to quench the remaining zero-point entropy in order to minimize the Zeeman energy further. Correspondingly, at the transition monopoles proliferate and are accommodated into the magnetic ground state at larger fields. 

Neglecting the magnetic Coulomb force among monopoles, this transition is approximated by a two-dimensional Ising model in a longitudinal field 
\cite{Isakov2004}. 
However, the long-range interaction promotes quantum fluctuations so that the transition becomes first-order and also extends to finite $T$ \cite{Castelnovo2008}. 
The resulting line of first-order transitions terminates at a critical endpoint  at a finite temperature $T_{\rm c} = 360$\,mK \cite{Sakakibara2003,Hiroi2003,Aoki2004,Kolland2012}. 
It was argued in Ref.~\cite{Shtyk2010} that the universality of this endpoint is equivalent to the one in critical uniaxial ferroelectrics
that show mean-field behavior with logarithmic corrections. 

In Kagome-ice, the low-energy monopole excitations carry an electric dipole that is parallel or antiparallel aligned with the applied magnetic field, see Fig.~\ref{fig:chi1-eps1}(a). 
Other configurations like Fig.~\ref{fig:chi1-eps1}(b) have an excitation energy of approximately 4\,K close to the critical field so that they are frozen out at lower temperatures. 
On the other hand, 
in the limit of large fields each tetrahedron houses a monopole whose electric dipoles are antiferroelectrically ordered with respect to each Kagome plane.
Correspondingly, it was suggested in Ref.~\cite{Khomskii2012} that antiferroelectricity is a secondary order parameter of the transition at $H_{i,c}$ with the concomitant signatures in $\varepsilon(\nu)$.

The magnetic field dependence of the magnetic response $\chi'(\nu)$ at $T=2$\,K is shown in the upper frame of Fig.~\ref{fig:chi1-eps1}(d) for frequencies up to 1\,kHz. 
At small fields $\chi'(\nu)$ is strongly dispersive but it becomes practically frequency independent at higher fields. 
This change of dispersion is less attributed to a field dependence of the relaxation rate but rather to a quench of the slow relaxation process as the magnetic field increases towards $H_{i,c}$, i.e. the weight of the loss peak in $\chi''(\nu)$ is suppressed with increasing $H_{i}$ \cite{Matthews2012,Bovo2013}.
The field dependence of the dielectric response to an ac electric field longitudinal to the [111] axis is displayed in the lower frame of Fig.~\ref{fig:chi1-eps1}(d). 
Again, $\varepsilon(\nu)$ is insensitive to the slow relaxation dominating $\chi(\nu)$ but only exhibits faster relaxation that becomes visible for frequencies exceeding 1\,kHz. 
At the same time, a pronounced magnetic field dependence appears with a maximum close to the critical field.

 \begin{figure}
 \centering
\includegraphics[width=\columnwidth]{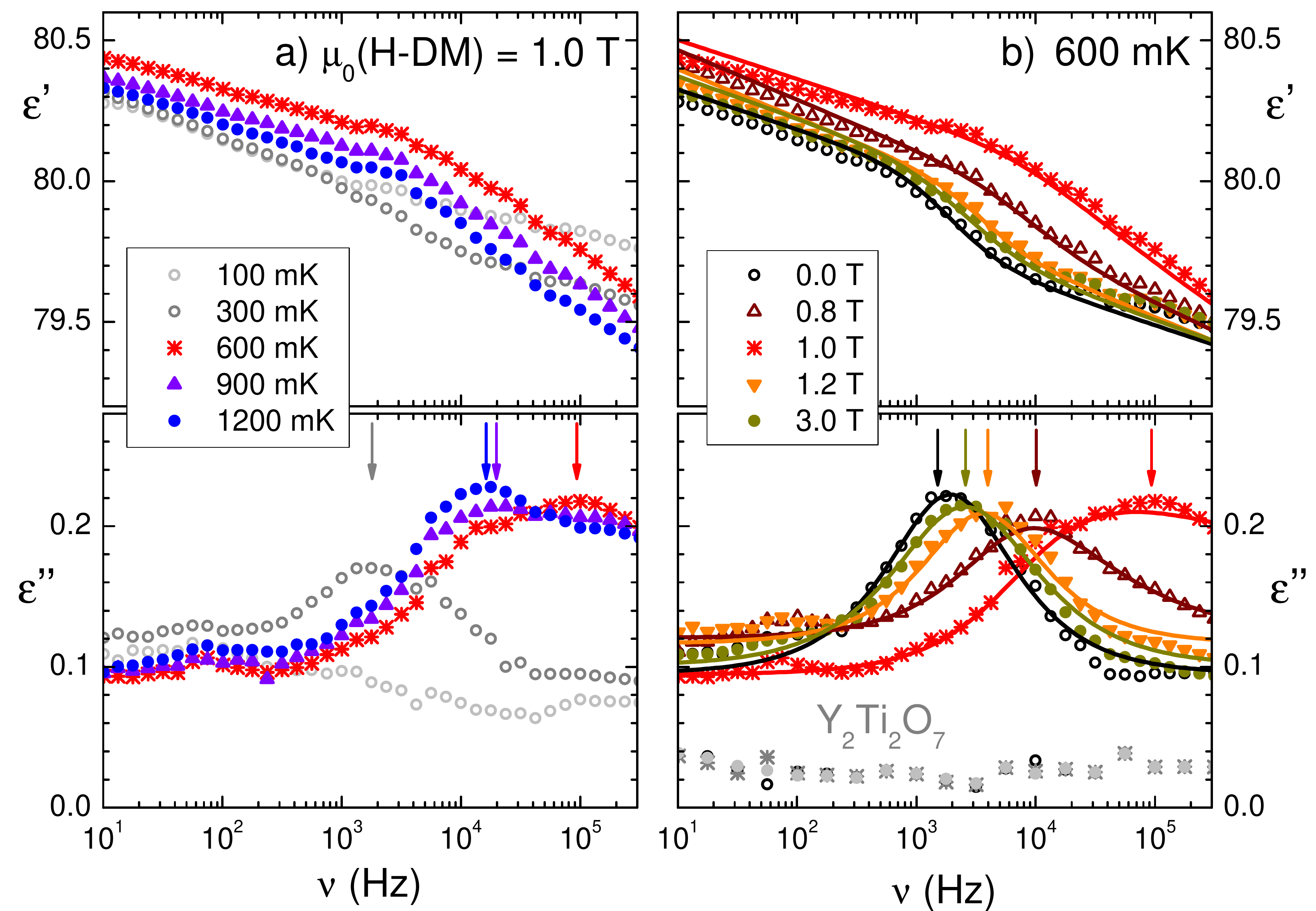}
 \caption{Full spectra of the dielectric response in \dto for (a) selected temperatures close to the critical field and (b) selected internal magnetic fields $H_{i}$ at $T= 600$\,mK.
 The step-like features in the real part $\varepsilon'$ and the corresponding loss peaks in the imaginary part $\varepsilon''$ shift with $T$ and $H_{i}$. 
 A maximal peak frequency $\nu_p \approx 100$\,kHz is obtained in the immediate vicinity of the critical endpoint. 
 The solid lines in (b) are fits to the Havriliak-Negami relaxation function \cite{Kremer2003} together with a background and constant loss contributions as described in the text.} 
 \label{fig:spectra}
\end{figure}

Full spectra of the dielectric response are presented in Fig.~\ref{fig:spectra} for selected magnetic fields and temperatures. 
The spectra contain a constant field-independent contribution to $\varepsilon''$, which translates via the Kramers-Kronig relations to $\varepsilon'_{\rm c.l.}(\nu) \sim -\ln{\nu}$ giving rise to a constant negative slope in the permittivity on a logarithmic frequency scale. 
Such a constant loss is known from other titanate based quantum-paraelectric materials like SrTiO$_3$ \cite{Hemberger1994}. 
It has been attributed to quantum tunneling between narrow off-centered positions of the Ti-ions, which are also responsible for the very high absolute value of the low-temperature permittivity of about $\varepsilon_\infty\approx 80$. 
A comparison of the dielectric loss with measurements on the non-magnetic iso-structural compound \yto at 600\,mK plotted in grey in Fig.~\ref{fig:spectra}(b) shows that the peak is absent while a constant loss contribution remains.
This constant loss was subtracted from the data shown in Fig.~\ref{fig:chi1-eps1}.

After considering 
the large constant background and the constant loss, the remaining dielectric relaxation was modeled using a Havriliak-Negami function, which in principle can be related to an asymmetrically broadened distribution of relaxation times (see \cite{Kremer2003} and SI).
It yields a very good description of the 
complex permittivity spectra as demonstrated in Fig.~\ref{fig:spectra}(b). 
A full set of fitting parameters is given in the SI. 

\begin{figure}
 \centering
 \includegraphics[width=1\columnwidth]{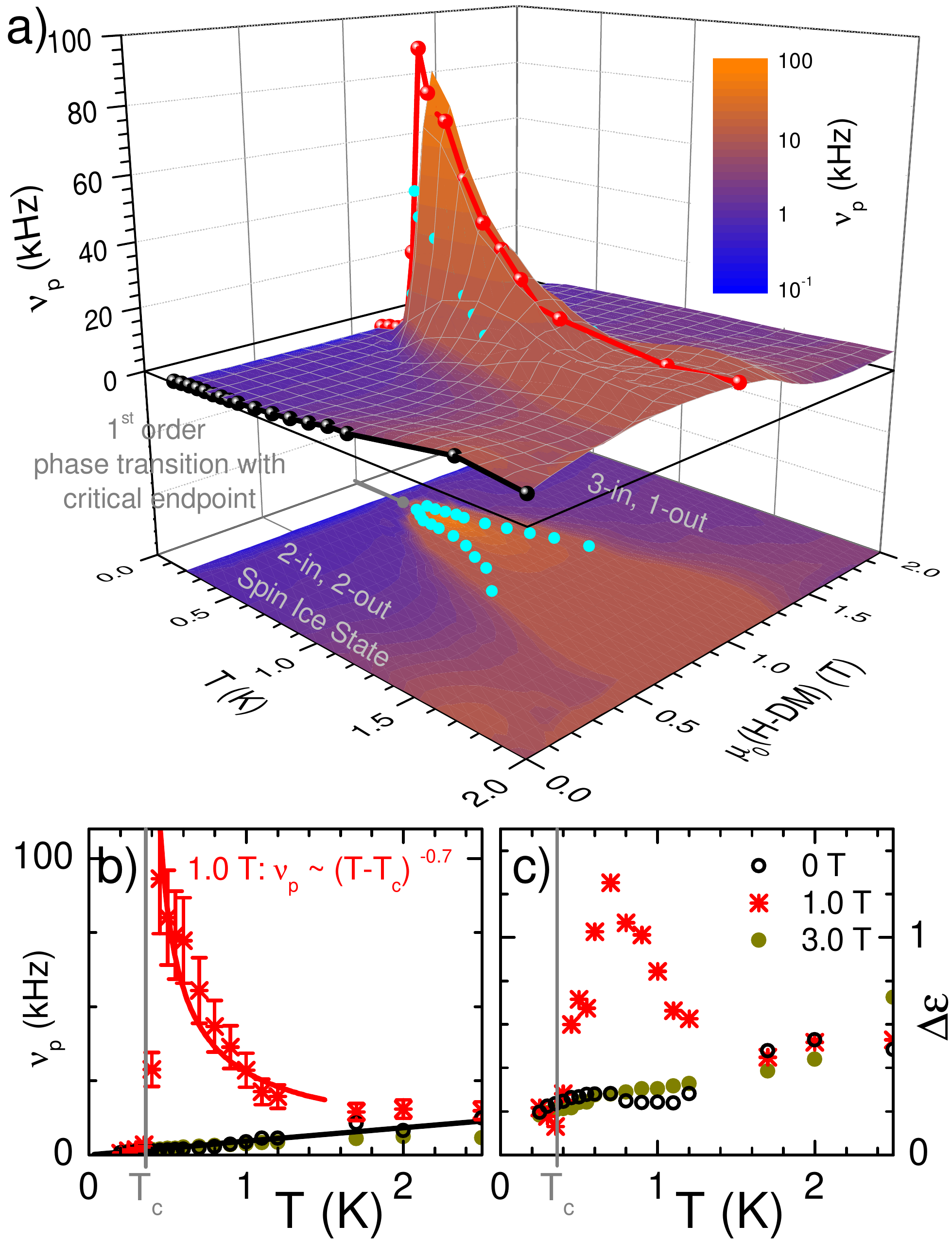}
 \caption{(a) Peak frequency $\nu_p = 1/(2\pi\tau)$ of the dielectric loss spectra as a function of temperature, $T$, and 
 internal field $H_i = H-D M$ along [111] axis.
 %
 The critical endpoint temperature is $T_{\rm c} = 360$\,mK. 
 The black and red curve mark $\nu_p(T)$ at zero and critical field; the blue dots mark the positions of FWHM of the $\nu_p(H_i)$ curves at different temperatures. (b) Temperature dependence of the relaxation frequency $\nu_p$ and (c) of the relaxation strength $\Delta\varepsilon$ at three different internal magnetic fields.}
 \label{fig:tau-contour3d}
\end{figure}

The most important parameter characterizing the relaxational dynamics is the position, $\nu_p$, of the loss peak (see arrows in Fig.~\ref{fig:spectra}) that defines a dielectric relaxation rate $1/\tau = 2\pi \nu_p$.
The dependences of $\nu_p(T, H_i)$ for temperatures, $T$, below 2\,K and internal magnetic fields, $\mu_0 H_i$, up to 2\,T 
is shown in Fig.~\ref{fig:tau-contour3d}(a). 
Away from the critical field, $\mu_0 H_{i,c} = 1$ T,
the relaxation rate is remarkably independent of field.
As shown in more detail in Fig.~\ref{fig:tau-contour3d}(b), the rate  at zero field and 3\,T almost coincide.
Such a field-independent relaxation in the dielectric response is likely to be of non-magnetic origin although the comparison in Fig.~\ref{fig:spectra}(b) with the non-magnetic compound \yto suggests otherwise. 
However, a strong magnetic influence indisputably appears close to the critical field.
The relaxation rate changes dramatically as a function of field and temperature, and it reflects the critical behavior associated with the monopole liquid-gas endpoint.
We therefore conclude that the dielectric dynamics is sensitive to the criticality of magnetic monopoles close to their condensation transition.

Besides establishing the 
magnetoelectric coupling in spin-ice, the results of Fig.~\ref{fig:tau-contour3d} also reveal unexpected critical dynamics. 
Away from the critical field, the relaxation rate decreases monotonically with temperature, see SI for more details.
%
%
However, as the critical endpoint is approached, the relaxation changes dramatically with a sharp increase in $\nu_p$. 
This implies that the polarization decorrelates faster close to criticality.
Assuming critical dynamic scaling, we can relate the relaxation time to the correlation length, $\tau \sim \xi^z$, via the dynamical exponent $z$.
At the critical field the correlation length is assumed to diverge as a function of temperature, $\xi \sim |T-T_{\rm c}|^{-\nu_\xi}$ with the correlation length exponent $\nu_\xi > 0$, so that 
\begin{equation}
\frac{1}{\tau(T)}\Big|_{H_i=H_{i,c}} = 2\pi \nu_p\Big|_{H=H_{i,c}} \sim |T-T_{\rm c}|^{\nu_\xi z}.
\end{equation}
Usually, the dynamical exponent $z$ is positive resulting in the canonical behavior of critical slowing-down.
Here, we find instead a {\it negative} dynamical exponent implying a {\it critical speeding-up} of the dynamics.
Close to the critical temperature this speeding-up is particularly dramatic as a function of field with $\nu_p(H_i)$ growing by two orders of magnitude.
From a fit within the temperature range from 450\,mK to 1.2\,K we can estimate the exponents to be approximately $\nu_\xi z \approx -0.7$, see Fig.~\ref{fig:tau-contour3d}(b).

\begin{figure}
 \centering
 \includegraphics[width=1\columnwidth]{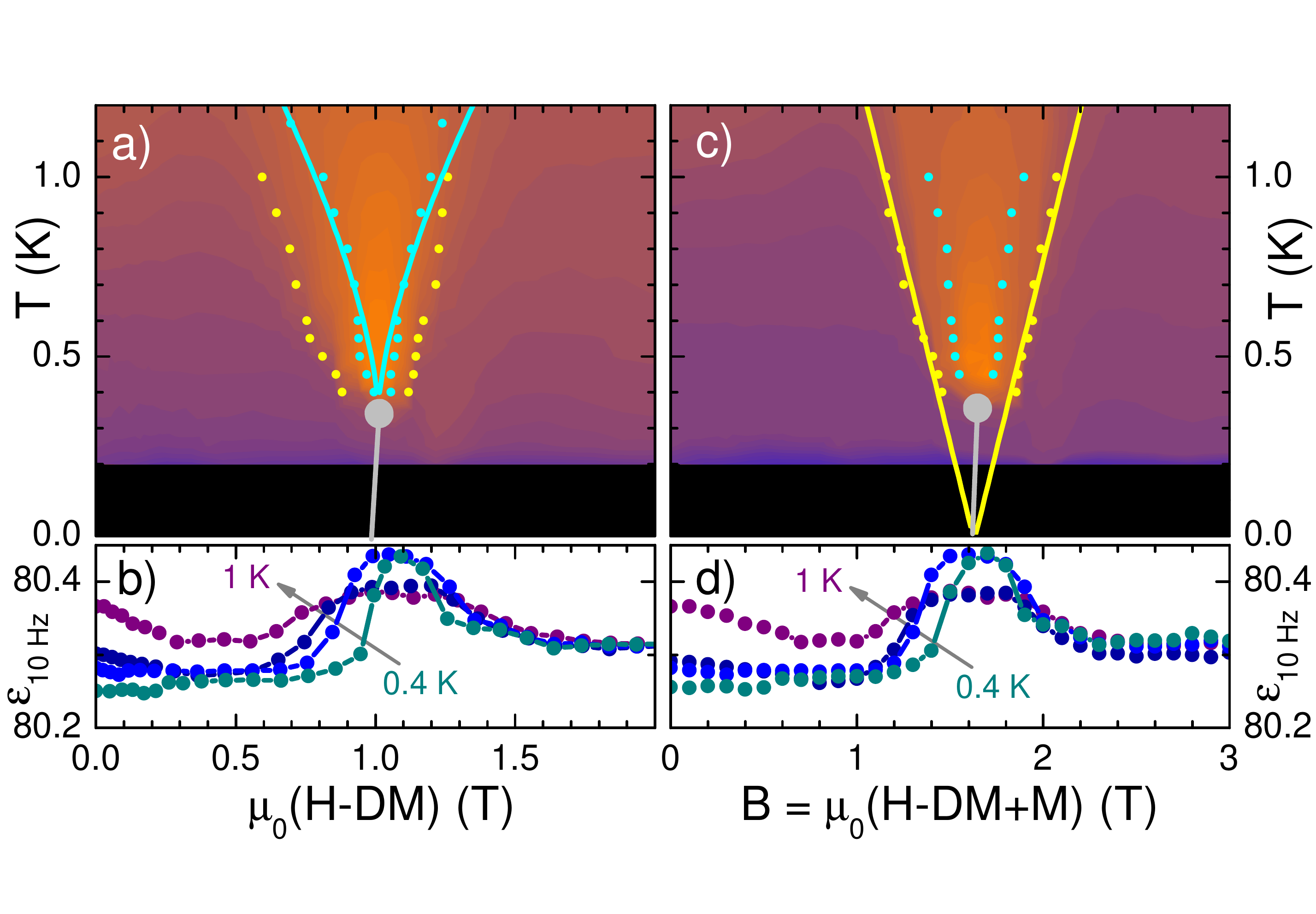}
 \caption{(a) $(H_i,T)$-diagram and contour plot of the peak frequency $\nu_p = 1/(2\pi\tau)$ of the dielectric loss  (same colour legend as in Fig.~\ref{fig:tau-contour3d}(a)). 
The position of FWHM of $\nu_p(H_i)$ (blue dots) approximately obey $T-T_{\rm c} \sim |H_i - H_{i, c}|^{2/3}$ scaling (blue line).
(b) Quasi-static permittivity $\varepsilon_{\rm 10\,Hz}$ measured at $\nu = 10$\,Hz.
The inflection points of its field dependence are marked in (a) as yellow dots. (c) Same data as in panel (a) but plotted versus the magnetic flux density $B = \mu_0 (H_i + M)$. Here, the locations of the yellow dots linearly extrapolate to zero temperature (yellow lines).
 }   
 \label{fig:figure4}
\end{figure}

We can define a crossover scale by locating the positions of the full width at half maximum (FWHM) of the function $\nu_p(H_i)$, which yield the blue dots in Fig.~\ref{fig:tau-contour3d}(a). 
These crossover lines first follow the mean-field prediction \cite{Shtyk2010}, $T-T_{\rm c} \sim |H_i-H_{i,c}|^{2/3}$, as shown by the blue solid lines in Fig.~\ref{fig:figure4}(a). 
A similar scaling close to the endpoint is also observed in the 
specific heat \cite{Hiroi2003}.
Closer inspection, however, reveals deviations very close to the critical temperature. 
Indeed, critical speeding-up comes to an halt at a temperature of 450\,mK just before $T_{\rm c} = 360$\,mK, and $\nu_p(T)$ abruptly drops towards zero, see Fig.~\ref{fig:tau-contour3d}(b)
This sudden qualitative change of behavior could have different origins. 
It might be attributed either to inhomogeneous demagnetization fields smearing the critical behavior or a slight misalignment of the applied field with respect to the [111] axis. 
Alternatively, it could indicate a crossover to a more conventional behavior of critical slowing-down.
The relaxation process is also very asymmetric with respect to the critical temperature.
It is indistinguishable from the non-critical background for $T < T_{\rm c}$.
 
The weight of the dielectric loss, i.e., the relaxation strength $\Delta \varepsilon$ is shown in Fig.~\ref{fig:tau-contour3d}(c).
Again, the weight at zero field and 3\,T are barely distinguishable but a strong enhancement appears close to the critical field.
As a function of decreasing temperature, the weight starts to increase at around 1.5\,K but peaks at approximately 700\,mK before dropping back to the non-critical background.
The location of this peak position emphasizes the qualitative change of the relaxation dynamics upon approaching the critical temperature as already observed in the relaxation rate. 

The quasi-static limit of the dielectric response, $\varepsilon_{\rm 10\,Hz}$, measured at $\nu = 10$\,Hz is shown in Fig.~\ref{fig:figure4}(b). As pointed out in Ref.~\cite{Saito2005}, the two plateaus emerging at low temperatures can be related to a magnetic renormalization of the permittivity, $\delta \varepsilon(0) \sim \langle M^2\rangle$, and reflect the constant magnetization of Kagome ice and its saturation at low and high fields, respectively.
In principle, one expects that the static permittivity exhibits signatures at the critical crossover, i.e., at the blue lines Fig.~\ref{fig:figure4}(a). 
However, such signatures are not observed and are possibly beyond the resolution of our measurement.
The peak of $\varepsilon_{\rm 10\,Hz}$ close to the critical field gives rise to inflection points whose positions are marked as yellow dots in Fig.~\ref{fig:figure4}(a),
which instead define an additional crossover that, interestingly, is markedly different from the one extracted from the FWHM of $\nu_p$. 
It is characterized by a different $T$-dependence and and it is also asymmetric with respect to the critical field. 
The presence of an additional crossover and its asymmetry is unexpected for an endpoint in the Ising universality class. 

The same data is plotted in Fig.~\ref{fig:figure4}(c) against the magnetic flux density $B = \mu_0 (H_i + M) \approx \mu_0 H$ that approximately corresponds to the applied magnetic field for our disc-shape sample, $D \approx 1$. Here, remarkably, the additional crossover becomes symmetric and extrapolates linearly to zero temperature (yellow lines), $T \sim |B-B_c|$. Such a linear scaling to zero temperature is reminiscent of the effective short-range Ising model for spin-ice \cite{Isakov2004}. 
We therefore interpret tentatively the yellow crossover extracted from the quasi-static permittivity as a signature for the onset of critical short-range correlations. Interestingly, the development of these correlations coincides with the magnetic field dependence of the dielectric dynamics. Outside the regime enclosed by the yellow crossover lines the dielectric relaxation is practically field independent but it gets strongly modified inside indicated by its orange shading in Fig.~\ref{fig:figure4}(a) and (c).

The emergent magnetic monopoles in spin-ice are accompanied by an electric dipole moment giving rise to intrinsic magnetoelectric coupling \cite{Khomskii2012}.
We confirmed this magnetoelectricity in \dto and, in addition, demonstrated that the dielectric dynamics gets dramatically modified close to the monopole liquid-gas transition resulting in a critical speeding-up. 
Besides revealing a peculiar aspect of the dynamical critical behavior of monopole condensation, 
our findings provide a proof of principle for dielectric  spectroscopy of monopoles in spin-ice.

\acknowledgements
We thank S. Bramwell, D. Khomskii, T. Lorenz, and A. Rosch for useful conversations. This work is supported by the Deutsche Forschungsgemeinschaft via SFB608 (Cologne), FOR960, and research grant HE-3219/2-1.

\clearpage 
\newpage
\section{Supplementary information}

\subsection{Methods}

Single crystals of \dto were grown using a floating zone method. 
The structural characterization by X-ray diffraction revealed phase purity and the expected cubic structure.
The samples were prepared as thin platelets normal to the [111] direction with a typical thickness of $d\approx 0.5$\,mm and an area of $A\approx 10$\,mm$^2$.
The measurements of the complex dielectric permittivity were performed in the frequency range from 1\,Hz to 1\,MHz employing a frequency-response analyzer {\sc Novocontrol Alpha-A} with stimuli of typically 1\,V/mm. 
For these measurements contacts of silver paste were applied at opposite sides of the crystal platelets. 
The samples were oriented with magnetic and electric fields in the crystalline [111] direction.
The magnetic-field range up to 3.5\,T and the temperature range from 100\,mK to 5\,K for the dielectric measurements were covered using a {\sc QuantumDesign PPMS} magneto-cryostat and a {\sc Oxford Kelvinox} $^3$He/$^4$He dilution refrigerator. 
The magnetic AC-susceptibility measurements were performed in a {\sc QuantumDesign MPMS} SQUID with AC-Option with AC-magnetic field stimuli of 2\,Oe parallel to the applied DC-field along [111] direction.

\subsection{Constant loss behaviour in quantum paralectric titanates}

Titanates with Ti-$3d^0$ configuration in an octahedral oxygen environment usually show a large polarizability due to the off-center motion of the titanium ions. 
Aside from the prototypical ferroelectric BaTiO$_3$ the quantum paraelectric sister compound  SrTiO$_3$ is a well known example \cite{Hemberger1994}.
In quantum paraelectrics the ferrolectric order is suppressed due to quantum fluctuations, leading to a saturation of the dielectric permittivity at relatively large values at low temperatures. 

Within this regime the dielectric response exhibits a constant, i.e.\ frequency independent, loss contribution \cite{Hemberger1994}.
Such a contribution in the imaginary part of the response function $\varepsilon''_\text{c.l.}$ is via the Kramers-Kronig relations connected to a corresponding contribution in the real part, which shows a constant negative slope: 

\begin{equation} \label{eq:CL}
  \varepsilon'(\nu) =  - \frac{2\varepsilon''_\text{c.l.}}{\pi}\ln{\frac{\nu}{\nu_0}} + \varepsilon_\infty
\end{equation}
\begin{equation}
  \varepsilon''(\nu) =  + \varepsilon''_\text{c.l.} 
\end{equation}

\begin{figure}
 \centering
 \includegraphics[width=0.7\columnwidth]{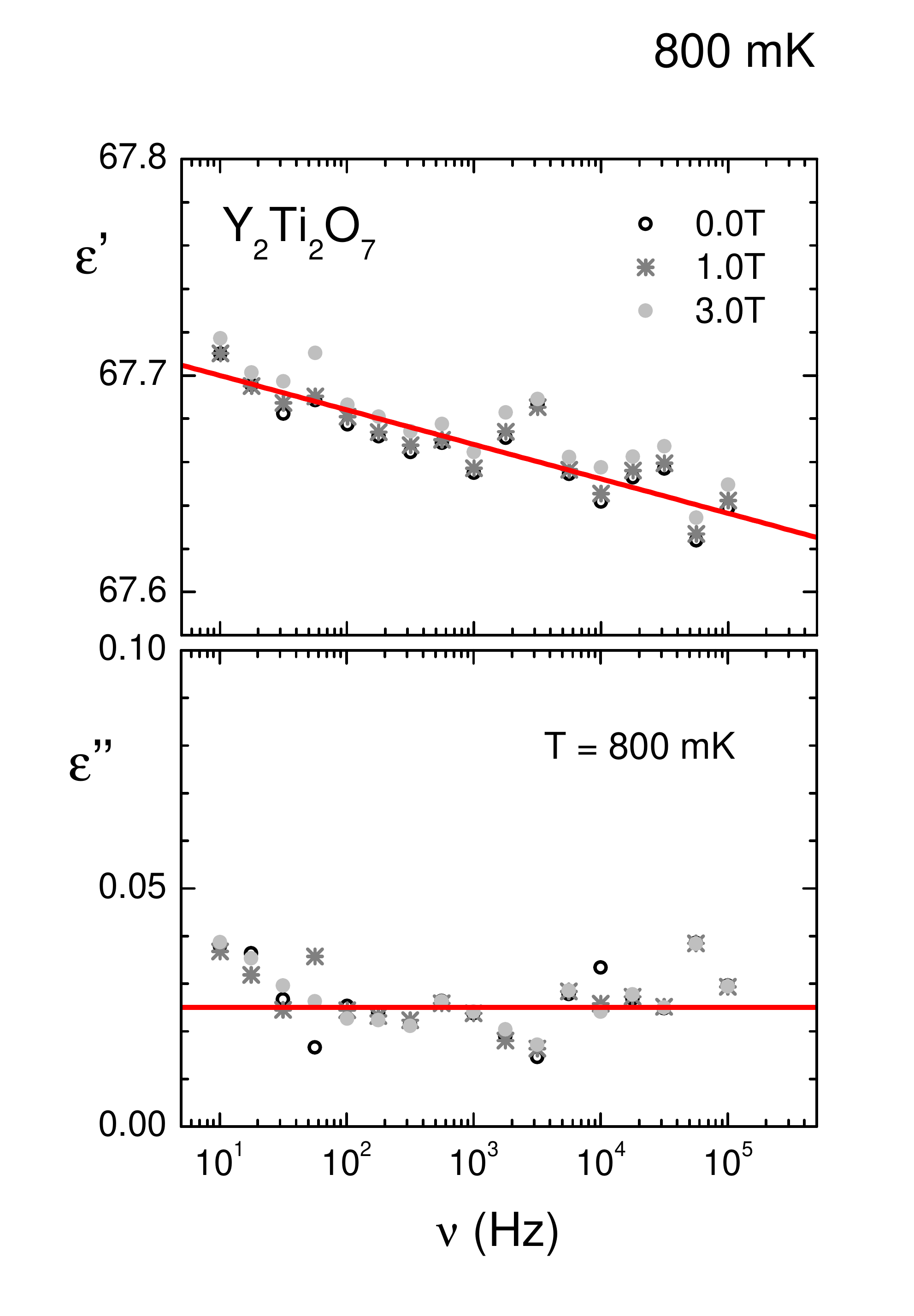}
 \caption{Complex permittivity of \yto measured at 800\,mK for various magnetic fields. The solid lines are calculated using Eq.~\ref{eq:CL}.}
 \label{fig:yto-cl}
\end{figure}

Here $\nu_0$ denotes a reference frequency which in the following was chosen as 10\,Hz. 
In Fig.~\ref{fig:yto-cl} we show measurements of the dielectric response of non-magnetic \yto.
On top of the typically large dielectric background of $\varepsilon_\infty$ one finds the signatures of constant loss behavior, i.e.\ a frequency independent contribution to $\varepsilon''(\nu)$ and a negative slope on $\varepsilon'(\nu)$.  
Such a type of contribution can also be found in \dto as described in the paper.

\subsection{The Havriliak-Negami function}

For the evaluation of the peak position, the step height $\Delta\varepsilon$ in $\varepsilon'$ and the relaxation time $\tau$, the spectra of the dielectric loss were fitted with a Havriliak-Negami function \cite{Kremer2003} and a nearly constant loss contribution:

\begin{figure}
 \centering
 \includegraphics[width=1\columnwidth]{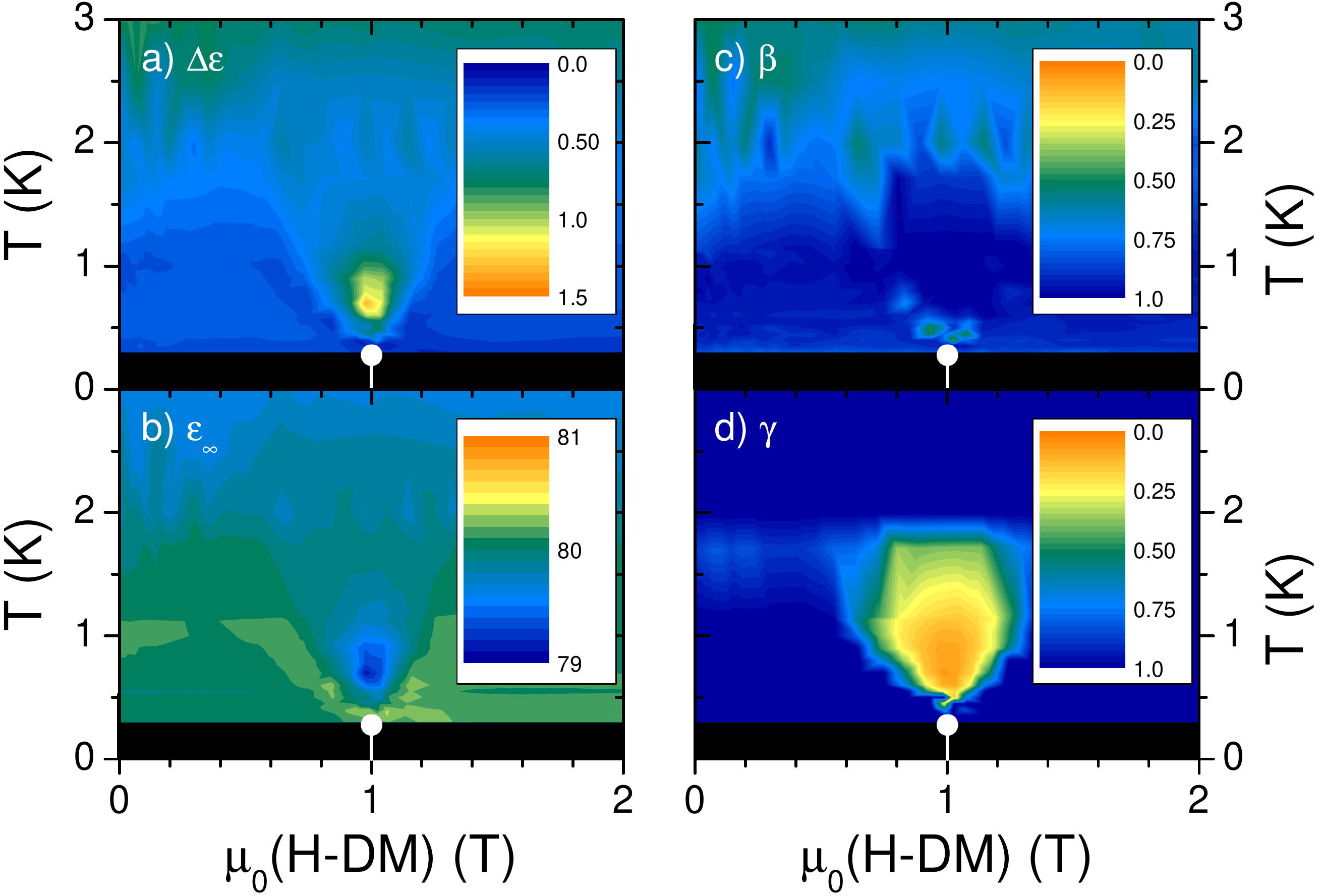}
 \caption{Temperature and magnetic field dependence of fitting parameters above the critical endpoint according to Eq.~(\ref{eq:HN}). 
The contour plots are interpolations of magnetic field sweeps at different temperatures.
(a) Relaxation strength $\Delta \varepsilon$, (b) high-frequency limit $\varepsilon_{\infty}$, (c) broadening parameter $\beta$ and (d) asymmetry parameter $\gamma$.}
  \label{fig:HN-parameters} 
\end{figure}

\begin{equation} \label{eq:HN}
  \varepsilon'(\nu) = \Delta\varepsilon \cdot \text{Re}\left[\frac{1}{\left(1+\left(i2\pi\nu\tau_\text{HN}\right)^\beta\right)^\gamma}\right] 
  - \frac{2\varepsilon''_\text{c.l.}}{\pi}\ln{\frac{\nu}{\nu_0}} + \varepsilon_\infty 
\end{equation}
\begin{equation}
  \varepsilon''(\nu) = \Delta\varepsilon \cdot \text{Im}\left[\frac{1}{\left(1+\left(i2\pi\nu\tau_\text{HN}\right)^\beta\right)^\gamma}\right] + \varepsilon''_\text{c.l.} 
\end{equation}

The shape of the permittivity spectra is determined by the parameters $\beta$ and $\gamma$ ranging between unity and zero.
For $\beta=\gamma=1$ the Havriliak-Negami function replicates the Debye-relaxation.
If $\beta < 1$ the spectra $\varepsilon''(\log\nu)$ are symmetrically broadened with respect to the Debye case.
For $\gamma < 1$ the spectra are asymmetrically broadened. 
Both cases reflect the presence of a (symmetric or asymmetric)  distribution of relaxation times in contrast to the mono-dispersive Debye-relaxator.
In that case the parameter $\tau_\text{HN}$ does not reflect the most probable relaxation time $\tau$, i.e.\ the inverse of the peak frequency $2\pi\nu_\text{p}$.
But the fit parameters can be used to calculate the peak position $\nu_\text{p}$ and the corresponding relaxation time $\tau$ using

\begin{equation}
  \frac{1}{\tau} = 2\pi\nu_\text{p} = \frac{1}{\tau_\text{HN}}\left[\sin\frac{\beta\pi}{2+2\gamma}\right]^{1/\beta}\left[\sin\frac{\beta\gamma\pi}{2+2\gamma}\right]^{-1/\beta}
  \label{eq:HN-tau}
\end{equation}

Fig.~\ref{fig:HN-parameters} displays the results for the parameters $\Delta\varepsilon$, $\varepsilon_\infty$, $\beta$, and $\gamma$. 
Whereas the broadening parameter $\beta$ in panel (c) stays close to unity within the examined $(T,H)$-regime, the other fitting parameters vary substantially. In particular, it is apparent that at $H_{i,c}$ the fitting parameters assume maximal values approximately $0.5$ K above the critical temperature $T_c = 360$ mK. This indicates a qualitative change of the relaxation dynamics as the critical endpoint is approached, see also discussion in the main text.


\subsection{Demagnetization fields}

For technical reasons, dielectric measurements were performed on a disc-like sample of \dto with a demagnetization factor $D$ that is almost 1 resulting in substantial demagnetization fields within the sample.
The internal magnetic field strength $H-DM$ has been calculated numerically using magnetization data for \dto in [111] magnetic field published in \cite{SI-kolland}, the radia software package provided by the ESRF \cite{SI-radia}, and the dimensions of the rectangular sample.

\subsection{Temperature dependence of dielectric relaxation away from the critical field}

\begin{figure}[b]
 \centering
 \includegraphics[width=1\columnwidth]{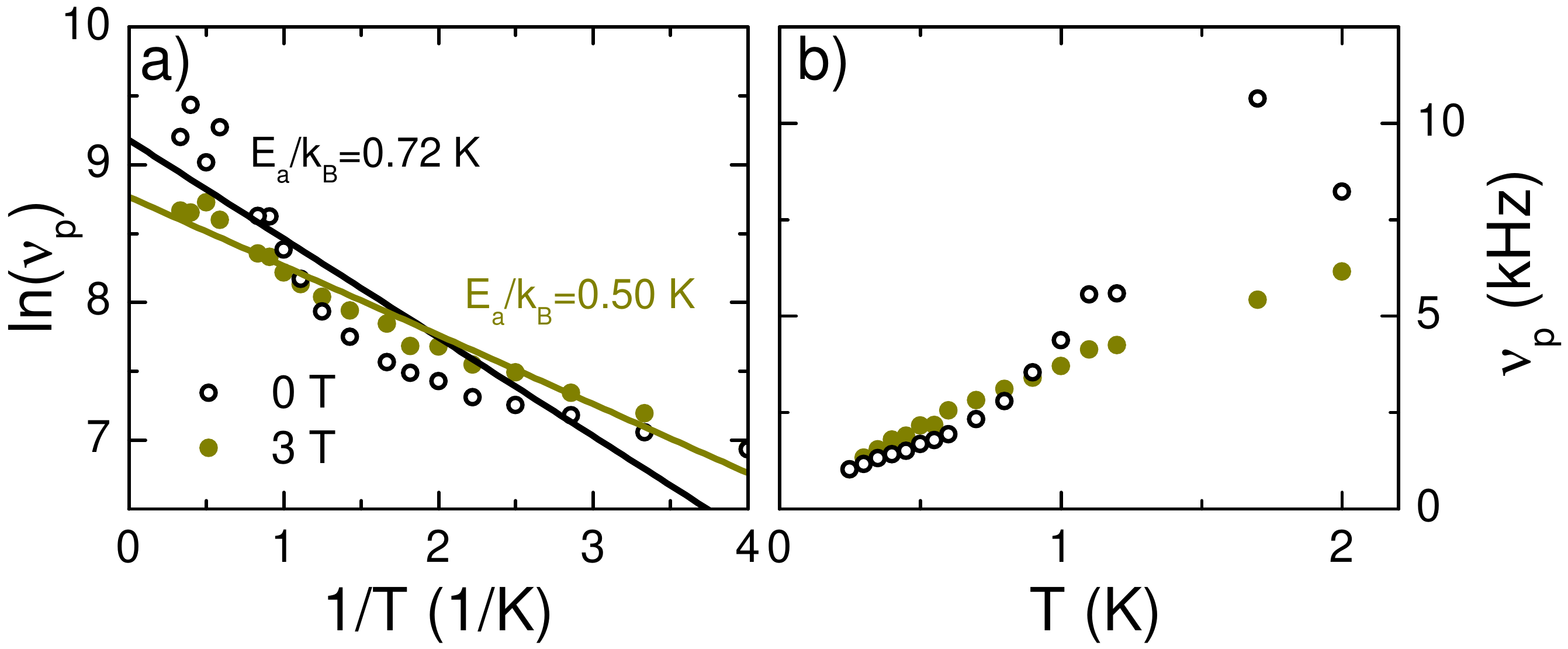}
 \caption{Temperature dependence of the peak frequency $\nu_p$ of the dielectric loss (a) on a logarithmic and (b) on a linear scale at zero field and $3$ T.
 The lines in (a) are fits to an Arrhenius law $\nu_p \sim $ exp$[-E_a/k_B T]$ with an activation energy $E_a$.
  }
  \label{fig:arrhenius} 
\end{figure}

Away from the critical field, the dielectric relaxation is approximately independent of magnetic field, see Fig.~\ref{fig:figure4}(c) of the main text. Fig.~\ref{fig:arrhenius} focusses on the temperature dependence of the peak frequency $\nu_p$ at zero field and $3$ T.
Both decrease monotonically with temperature but the precise functional $T$ dependence is not evident. At $3$ T the dependence can be approximately described by an 
Arrhenius law with an activation energy of $E_a/k_B \approx 0.5$ K, see solid lines.

\end{document}